\documentclass[superscriptaddress,twocolumn]{revtex4-1}

\usepackage{graphicx}
\usepackage{dcolumn}
\usepackage{bm}

\usepackage[latin1]{inputenc}
\usepackage{amsmath}
\usepackage{multirow}
\usepackage{ulem}

\begin{document}

\title{The effect of dissipation in direct communication scheme}
\author{Fu Li}%
\affiliation{%
Beijing Computational Science Research Center, Beijing 100084, P.R. China}
\author{Jun-Xiang Zhang}%
\affiliation{%
The State Key Laboratory of Quantum Optics and Quantum Optics Devices, Institute of Opto-Electronics, Shanxi University, Taiyuan 030006, China}
\author{Shi-Yao Zhu}%
\affiliation{%
Beijing Computational Science Research Center, Beijing 100084, P.R. China}
\affiliation{
Hefei National Laboratory for Physical Sciences at Microscale and Department of Modern Physics,
University of Science and Technology of China, Hefei, Anhui 230026, China}
 \date{\today}
\begin{abstract}
The effect of the dissipation and finite number of beam splitters are discussed. A method using balanced dissipation to improve the communication for finite beam splitters, which greatly increases communication reliability with an expense of decreasing communication efficiency.
\end{abstract}
\maketitle

\section{Introduction}
Quantum information is a rapidly developing area in recent decades. One of most important applications in quantum information is quantum communication. In 1984, Bennett and Brassard proposed the famous protocol for quantum key distribution (QKD), known as BB84[1], which is the first practical quantum information processor[2,3]. In the BB84 protocol, the security of protocol is guaranteed using single photon sent by Alice. For a high loss or not a perfect single photon state, Hwang proposed a decoy-pulse method to guarantee the security under the photon-number-splitting attack for the BB84[4]. Recently, A. Rubenok proposed a new quantum-key-distribution protocol that is immune to attack to vulnerabilities of single-photon detectors[5]. Another celebrated QKD protocol is E91[6], which is based on the quantum entanglement. At same time, using the electromagnetic field amplitudes of ``non-classical" light beams (squeezed or entangled light) to QKD also draw much attention [7-12]. Protocol of counterfactual quantum cryptography is proposed [13], and its experimental demonstration is reported [14]. Another protocol of counterfactual quantum cryptography is discussed with tripartite[15].

Recently, Salih et.al. proposed a protocol to realize direct counterfactual communication (no need for a prior quantum key distribution)[16] based on the previous work[17], which shown how to make an interaction-free measurement[18-20].  In[16], the `chained' quantum Zeno effect and interference of optical paths is used to achieve information transmission between Alice and Bob without any photon traveling between them, by using a single photon source. However, in order to have a direct communication, large number of perfect beam splitters (8000 pieces for 90\% efficiency[16]) and no dissipation in all paths are required. In a real experiment, the BSs are finite (usually less BSs is better) and the dissipation of the paths (including the BSs themselves) could not be avoided[21-23]. Here we consider the effect of the dissipation on the direct communication, and analyze how the reliability of the direct communication can be preserved. In Sec. II, we derive the equations shown the effect of the dissipation by using quantum operators, which is independent of the input states. In Sec. III, numerical calculation is carried out to show the effects of the dissipation and the finite number of BSs. In Sec. IV, we propose a method (balanced dissipation) to improve the communication for finite number of BSs. In Sec. V, we give the conclusion.

\section{The equations derivation}

Consider the setup shown in Fig.1(a), which is composed with two chains of BSs, the inner chain and the outer chain. The two chains are formed by (M-1) and (N-1) head-tail connected Mach-Zehnder interferometers (MZIs), respectively[13]. Alice and Bob can use the setup to have information communication with a single photon field (or a coherent field). The outer chain contains M beam splitters (BS in green color) with the same reflectivity $R = {\cos ^2}{\theta _M}$  with  ${\theta _M} = \pi /(2M)$ and (M-1) mirrors (black color in Fig.1(a)), which are in the hands of Alice. Each inner chain is formed by N BSs (in blue color) with the same reflectivity of  $R = {\cos ^2}{\theta _N}$ with  ${\theta _N} = \pi /2N$ and 2(N-1) mirrors. The reflectivity of all the mirrors is 100\% without the dissipation. The inner chain (Fig.1(b)) has two parts, the BSs (blue color) and half of its mirrors in the hand of Alice and the other half mirrors in the hand of Bob. From the blue color BSs to the mirrors in Bob's hands, the field needs to pass through the transmission channel, which is publically accessible. Bob can block the paths in his hands with inserting blocks (small red color rectangles). The two outputs of each inner chain go to the outer chain and detector $D_{3i}$, respectively. Alice sends out her field, Bob can choose to insert his blocks or not in the paths in his side, and Alice measures the counting (or intensities) received by the two detectors, $D_1$, and $D_2$. For no dissipation and infinite M and N, when Bob inserts his blocks, Alice will see $D_2$ click, and when Bob does not insert his blocks, Alice will find $D_1$ click. Hence, Alice will know whether Bob inserts the blocks or not.

In real experiment, the dissipation of the paths (including the mirror itself) could not be avoided. Here we group all the paths in Fig.1(a) into three groups, left (paths at the left line), middle (paths in the middle line) and right (paths in the right line). We assume the dissipation of each path in the left group is the same (${\kappa _1}$) and the dissipation of each path in the middle group is the same (${\kappa _2}$) and the dissipation of each path in the right group is the same (${\kappa _3}$). If Bob choose to insert blocks in his paths, we have ${\kappa _3} = 1$. Besides, the number of N and M is finite in real experiments. For finite N and M, the detection probability of $D_1$ (define as efficiency $W_1$) will not be 1 for no blocks, and the detection probability (efficiency) of $D_2$ ($W_2$)will not be 1 for with blocks, even no dissipation.

For finite M and N, in order to indicate how good the direct communication is, we introduce two quantities: 1. The probabilities (also called efficiencies), $W_1^{(nb)}$  (together with $W_2^{(nb)}$) for no blocks and $W_2^{(wb)}$ ($W_1^{(wb)}$) for with blocks, which represent the efficiency of the direct communication. Large $W_1^{(nb)}$ and $W_2^{(wb)}$ mean high efficient usage of the input photon for the communication. 2. The reliabilities, $\eta _{}^{(nb)} = W_1^{(nb)}/W_2^{(nb)}$ for no blocks and ${\eta ^{(wb)}} = W_2^{(wb)}{\text{/}}W_2^{(wb)}$ for with blocks, which represent how reliable the communication is. For no blocks, we want $W_2^{(nb)} \to 0$, while for with blocks, we want $W_1^{(wb)} \to 0$, so that Alice immediately knows that Bob inserts (or does not insert) his blocks when she sees the click of $D_2$ (or $D_1$). The larger ${\eta ^{(nb)}}$ and ${\eta ^{(wb)}}$ are, the more reliable the communication is.

\begin{figure}
\includegraphics[scale=0.5]{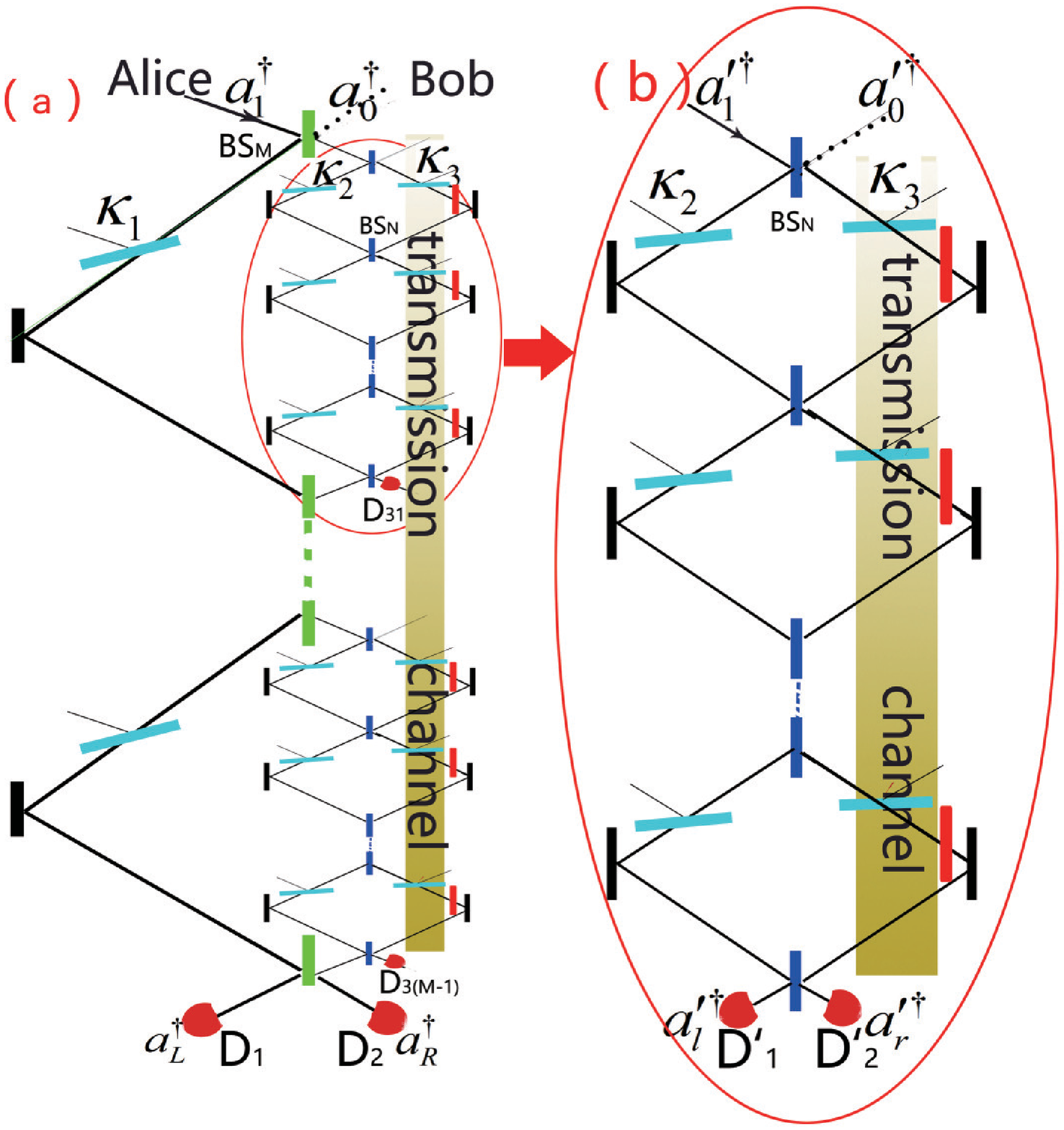}
  \caption{(color online)} \label{fig1}
\end{figure}

The dissipation in each path can be theoretically simulated by adding a beam splitter in the corresponding path, as shown in Fig.1 by the light cyan color, and the reflected energy is proportional to the intensity dissipation, ${\kappa _i}$ . The dissipation in a path can be theoretically calculated by multiplying a factor of  $\sqrt {1 - \kappa }$ to the path with   the dissipation of the path. Therefore, for example, the dissipation of one MZI in the inner chain can be represented by a matrix
 $\left[ {\begin{array}{*{20}{c}}
  {\sqrt {1 - {\kappa _2}} }&0 \\
  0&{\sqrt {1 - {\kappa _3}} }
\end{array}} \right]$.
The modes reflected by each added BS (for dissipation) is orthogonal to each other, and they are in vacuum for no input field. Therefore, we can treat all the reflected modes as one reservoir. The input field, $a_1^\dag $ , (together with two vacuum inputs ${a_0}^\dag$  and ${a'_0}^\dag$ ) is transformed into two outputs, $a_L^\dag {\text{ and }}a_R^\dag$  at $D_{1,2}$, the modes $a_{3i}^\dag$  at $D_{3i}$ and the reservoir modes $a_{res}^\dag$  (due to dissipation and blocks).

\begin{figure}
  \centering
  \includegraphics[scale=0.4]{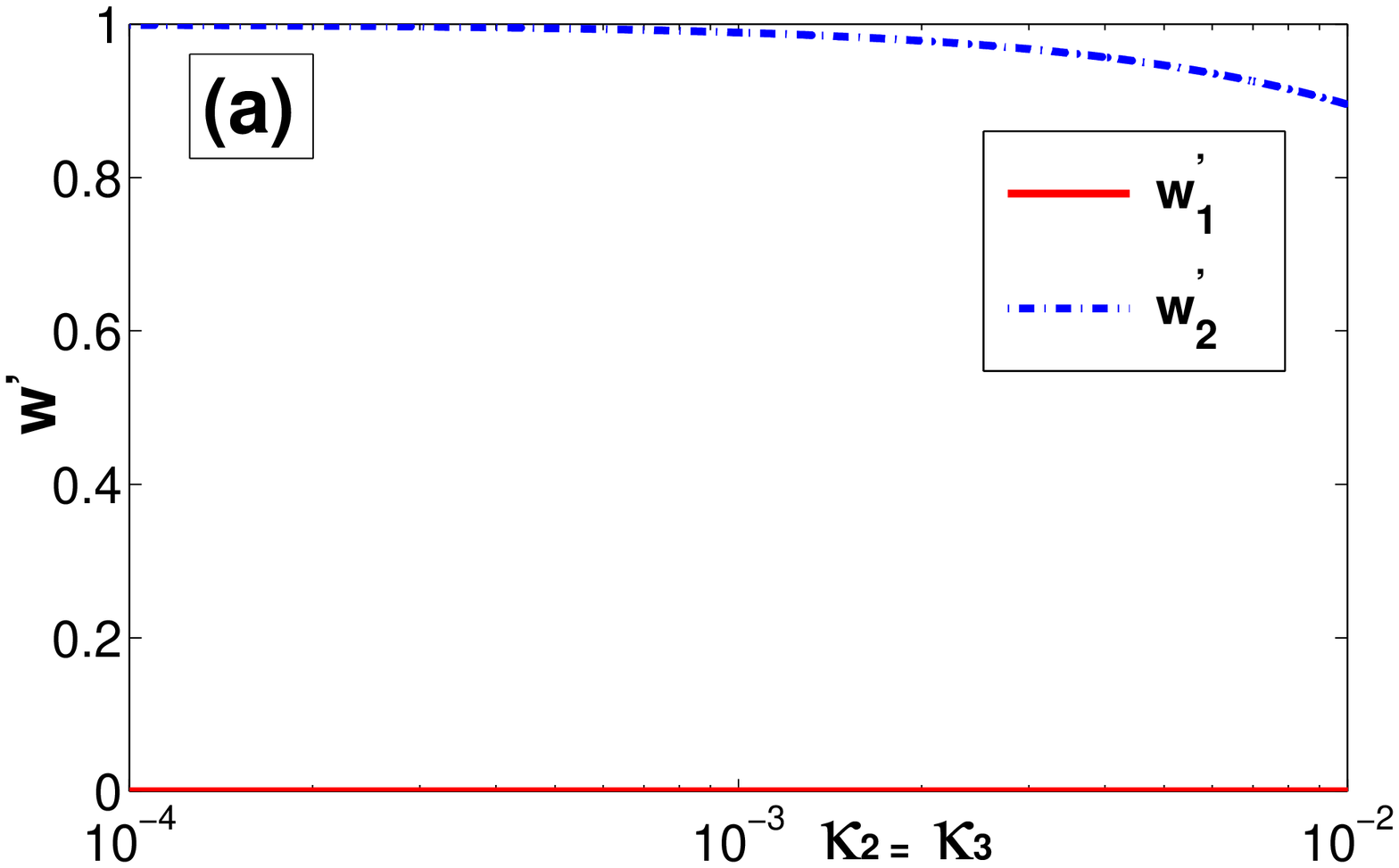}
   \includegraphics[scale=0.41]{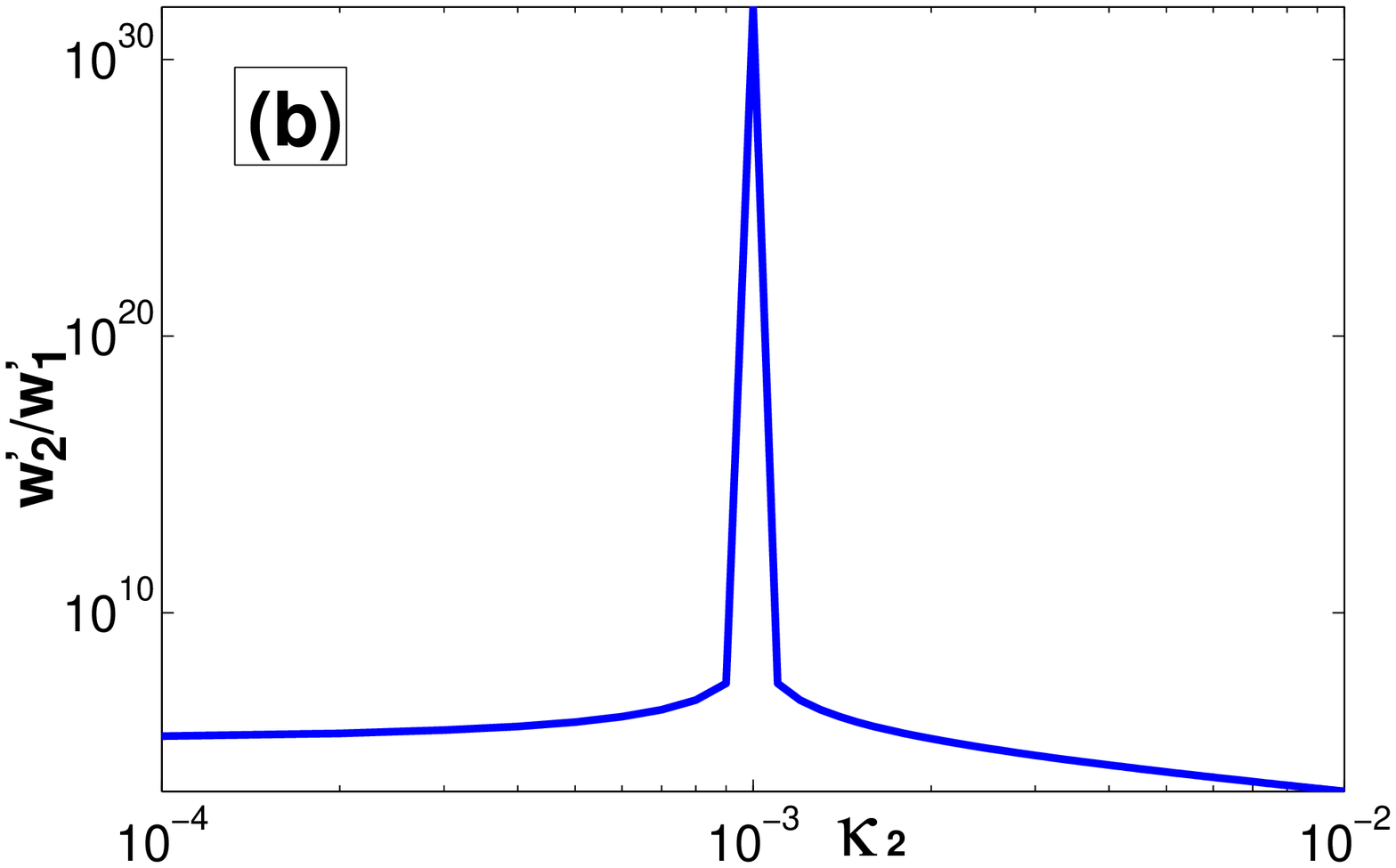}
\caption{The effect of the dissipation, (a) $W'_1$ and $W'_2$  versus  ${\kappa _3}$ with ${\kappa _2} = {\kappa _3}$ and N=12 (b) $W'_2/W'_1$  versus  ${\kappa _2}$ with ${\kappa _3} = {10^{ - 3}}$ and N=12  }\label{f1a}
\end{figure}

(i)  No blocks

Firstly, consider that Bob does not insert his blocks in the inner chain, and then the input will go to $D_1$, $D_2$ and $D_{3i}$, if there is no dissipation. The incident field of the inner chain (see Fig. 1(b)) is denoted by ${a'_1}^\dag $ , and the input vacuum is denoted by ${a'_0}^\dag $, and two outputs are denoted by ${a'_l}^\dag $  and ${a'_r}^\dag$ . Due to the dissipation, some photons go to the reservoir associated with the inner chain. Thus, the input ${a'_1}^\dag$  (the vacuum ${a'_0}^\dag$  has no contribution to the output) is transmitted into ${a'_l}^\dag $ ,  ${a'_r}^\dag$ and  ${a'_{res}}^\dag $ with,
\begin{equation}\label{inneroperator}
{a'_1}^\dag  \to {M'_{11}}{a'_l}^\dag  + {M'_{21}}{a'_r}^\dag  + {M'_{res}}{a'_{res}}^\dag
\end{equation}
where ${a'_{res}}^\dag $  is the creation operator of photons in the reservoir associated with the inner chain,${M'_{11}},{\text{ }}{M'_{21}}$ and ${M'_{res}}$  are transmission coefficients with${\left| {{{M'}_{11}}} \right|^2}{\text{ + }}{\left| {{{M'}_{21}}} \right|^2}{\text{ + }}{\left| {{{M'}_{res}}} \right|^2} = 1$ due to the photon number conservation. The two coefficients,  ${M'_{11}}$ and ${M'_{21}}$ are calculated in Appendix A,
\begin{subequations}
\begin{align} \label{innermatrix}
\small
{M'_{11}} =& [\begin{array}{*{20}{c}}1&0\end{array}]\left[ {\begin{array}{*{20}{c}} {\cos {\theta _N}}&{ - \sin {\theta _N}} \\ {\sin {\theta _N}}&{\cos {\theta _N}}\end{array}} \right] \\ \nonumber
                      &{\left[ {\begin{array}{*{20}{c}}
  {\sqrt {1 - {\kappa _2}} \cos {\theta _N}}&{ - \sqrt {1 - {\kappa _2}} \sin {\theta _N}} \\
  {\sqrt {1 - {\kappa _3}} \sin {\theta _N}}&{\sqrt {1 - {\kappa _3}} \cos {\theta _N}}
\end{array}} \right]^{N - 1}}\left[ {\begin{array}{*{20}{c}} 1 \\0\end{array}} \right]           \\
{M'_{21}} = &[\begin{array}{*{20}{c}} 0&1\end{array}]\left[ {\begin{array}{*{20}{c}}
  {\cos {\theta _N}}&{ - \sin {\theta _N}} \\{\sin {\theta _N}}&{\cos {\theta _N}}\end{array}} \right] \\   \nonumber
                      &{\left[ {\begin{array}{*{20}{c}} {\sqrt {1 - {\kappa _2}} \cos {\theta _N}}&{ - \sqrt {1 - {\kappa _2}} \sin {\theta _N}} \\
  {\sqrt {1 - {\kappa _3}} \sin {\theta _N}}&{\sqrt {1 - {\kappa _3}} \cos {\theta _N}} \end{array}} \right]^{N - 1}}\left[ {\begin{array}{*{20}{c}} 1 \\0\end{array}} \right]
\end{align}
\end{subequations}

The interference between the two paths of each MZI in the inner chain is important. Without dissipation, no-blocks will result in the complete interference, so that no photon enters ${D'_1}$  (photon completely entering  ${D'_2}$). If we have the same dissipation (balanced dissipation) for the two paths of each MZI in the inner chain,  ${\kappa _2} = {\kappa _3}$ (balanced dissipation), the complete interference  will be kept, and there is no photon entering  ${D'_1}$ , while the probability of the photon entering   ${D'_2}$ decreases with the dissipation${\kappa _2} = {\kappa _3}$ increasing, see Fig. 2a, where we plot the photon probabilities ($W'_1$ and $W'_2$) entering  $D'_1$  and   $D'_2$ versus ${\kappa _2}$ , (${\kappa _2} = {\kappa _3}$) with N=12, and no blocks. In Fig. 2b, we plot the probability ratio $W'_2/W'_1$  versus  ${\kappa _2}$ with ${\kappa _3} = {10^{- 3}}$, N=12 and no blocks. Please note the peak (infinity) at ${\kappa _2} = {\kappa _3}$  due to ${W'_1} = 0$ , because of the complete interference. For no blocks, with the balanced dissipation for each MZI of the inner chain (${\kappa _2} = {\kappa _3}$), the photon entering the inner chain will not return back to the outer chain ($W'_1 = 0$), which is independent of N (as the same of no dissipation). Therefore, the balanced dissipation in the inner chain will not affect the outer chain. That is to say, for ${\kappa _2} = {\kappa _3}$  and no block, the ratio of the photon probabilities (efficiencies) entering $D_1$ and $D_2$ is only determined by M, and are independent from the dissipation in the paths of the left group, ${\kappa _1}$, the same as the case of no dissipation. The ratio (also the reliability) is ${\eta ^{(nb)}}= {{{{\cos }^2}(\pi /2M)} \mathord{\left/
 {\vphantom {{{{\cos }^2}(\pi /2M)} {{{\sin }^2}(\pi /2M)}}} \right.
 \kern-\nulldelimiterspace} {{{\sin }^2}(\pi /2M)}}$ for no blocks and balanced dissipation,see Fig. 3(a).

(ii)  With Blocks

If Bob inserts his blocks, we need to consider the outer chain, which is also composed of many MZIs. The inner chain is one of the two paths of each MZI of the outer chain. The input field of the inner chain comes from the outer chain. The left outputs of the inner chain will go to the outer chain, and the right outputs (here indicated with) will be detected by Alice with detector $D_{3i}$, see Fig.1(a). The final outputs of the outer chain, $a_R^\dag $ and $a_L^\dag $ will be detected by $D_1$ and $D_2$ in Alice's hands. Using the method discussed above, we can get the total transformation for whole protocol, see appendix B.

\begin{equation}\label{final matrix}
  a_1^\dag  \to {M_1}a_L^\dag  + {M_2}a_R^\dag {\text{ + }}\sum\limits_i^{M - 1} {{M_{3i}}a_{3i}^\dag }  + {M_{res}}a_{res}^\dag
\end{equation}
where
\small
\begin{subequations}
\begin{align} \label{eqm1}
{M_1}= &[\begin{array}{*{20}{c}}1&0 \end{array}]\left[ {\begin{array}{*{20}{c}}{\cos {\theta _M}}&{ - \sin {\theta _M}} \\ {\sin {\theta _M}}&{\cos {\theta _M}}\end{array}} \right]    \\  \nonumber
              &{\left[ {\begin{array}{*{20}{c}}
  {\sqrt {1 - {\kappa _1}}\cos {\theta _M}}&-\sqrt {1 - {\kappa _1}}\sin {\theta _M} \\
  {{M'}_{11}\sin {\theta _M}}&{{M'}_{11}\cos {\theta _M}}
\end{array}} \right]}^{M - 1}\left[ {\begin{array}{*{20}{c}}
  1 \\
  0
\end{array}} \right]           \\
  {M_2} = &[\begin{array}{*{20}{c}} 0&1\end{array}]\left[ {\begin{array}{*{20}{c}}
  {\cos {\theta _M}}&{ - \sin {\theta _M}} \\
  {\sin {\theta _M}}&{\cos {\theta _M}}\end{array}} \right]  \\  \nonumber
                   &{\left[ {\begin{array}{*{20}{c}}
  {\sqrt {1 - {\kappa _1}}\cos {\theta _M}}&-\sqrt {1 - {\kappa _1}}\sin {\theta _M} \\
  {{M'}_{11}\sin {\theta _M}}&{{M'}_{11}\cos {\theta _M}}
\end{array}} \right]}^{M - 1}\left[ {\begin{array}{*{20}{c}}
  1 \\
  0
\end{array}} \right]
\end{align}
\end{subequations}

\normalsize
with (due to energy conservation)
\small
\begin{equation}\label{eqinitial}
{M_{res}} = \sqrt {1 - {{\left| {{M_1}} \right|}^2} - {{\left| {{M_2}} \right|}^2} - \sum\limits_{i = 1}^{M - 1} {{{\left| {{M_{3i}}} \right|}^2}} }
\end{equation}
\normalsize

In the above derivation, we use the operator transformation, while the input state is not specified. That is to say, we can use any input state for further derivation. We assume that the state of the input field can be written in the form of
\begin{equation}\label{singal photon}
\left| {{\psi _i}} \right\rangle  = f(a_1^\dag )\left| {\left\{ 0 \right\}} \right\rangle
\end{equation}
with $f(x)$ is an arbitrary function of the argument, $x$. Based on Eq.(3), the state of the output fields can be written as:
\begin{equation}\label{single output}
\small
\left| {{\psi _f}} \right\rangle  = f({M_1}a_L^\dag  + {M_2}a_R^\dag {\text{ + }}\sum\limits_i^{M - 1} {{M_{3i}}a_{3i}^\dag }  + {M_{res}}a_{res}^\dag )\left| {\left\{ 0 \right\}} \right\rangle
\end{equation}
\normalsize
If the input state is single photon, we have $f({a_1}^\dag ) = {a_1}^\dag$, and the final state is
\small
\begin{align}\label{eqfinial}
  \left| {{\psi _f}} \right\rangle &= {M_1}a_L^\dag  + {M_2}a_R^\dag {\text{ + }}\sum\limits_i^{M - 1} {{M_{3i}}a_{3i}^\dag }  + {M_{res}}a_{res}^\dag \left| {\left\{ 0 \right\}} \right\rangle \nonumber \\
                                                 &=\left| {{M_1},{M_2},{M_{31}}...{M_{3(M - 1)}},{M_{res}}} \right\rangle
\end{align}

For a coherent state input we have $f(a_1^\dag ) = {e^{ - {{\left| \alpha  \right|}^2}/2}}{e^{\alpha a_1^\dag }}$, the final state is
\small
\begin{align}\label{cofinial}
  \left| {{\psi _f}} \right\rangle &={e^{ - {{\left| \alpha  \right|}^2}/2}}{e^{\alpha ({M_1}a_L^\dag  + {M_2}a_R^\dag {\text{ + }}\sum\limits_i^{M - 1} {{M_{3i}}a_{3i}^\dag }  + {M_{res}}a_{res}^\dag )}}\left| {\left\{ 0 \right\}} \right\rangle  \nonumber \\
                                                 &= {\left| {{M_1}\alpha } \right\rangle _1}{\left| {{M_2}\alpha } \right\rangle _2}{\left| {{M_{31}}\alpha } \right\rangle _{31}}...{\text{ }}{\left| {{M_{3(M - 1)}}\alpha } \right\rangle _{3(M - 1)}}{\left| {{M_{res}}\alpha } \right\rangle _{res}}
\end{align}
\normalsize

In above,${\left| {{M_1}} \right|^2}$  ,${\left| {{M_2}} \right|^2}$ and ${\left| {{M_{3i}}} \right|^2}$ are the probabilities of a photon for the single photon input (or of intensities for the coherent state input) detected by $D_1$, $D_2$, and $D_{3i}$, respectively, and ${\left| {{M_{res}}} \right|^2}$ is the probability of a photon (or intensity) leaked out to the reservoir(some in Alice's hands and others in the transmission channel). Note that $D_1$, $D_2$, and $D_{3i}$ (i=1... M-1) are in the hands of Alice.
From Eqs. (8) and (9) we know that the proportions of field energy to each detector and the reservoir are independent of the input state. If we consider the ratios, we can get the same result with single photon or by using a coherent state. For single photon input, only one detector can have a click, or all detectors have no click if the one photon goes to the reservoir. If no click, we can throw away this communication (no information exchanged between Alice and Bob). Here we set the energy of the input state to be one. We discuss the proportions of energy entering each detector.
In the following diagrams, the energies received by detectors $D_1$, $D_2$, and $D_{3i}$ are labeled by $W_1$, $W_2$, $W_{3i}$, while the dissipated energy by $W_{res}$.
Here we would like to point out the difference between single photon input and coherent state input. For the single photon input, there is no photon in the transmission channel when $D_1$ or $D_2$ has a click (counterfactual), while for the coherent state input, there are photons in the transmission channel (no counterfactual). For the two different inputs, the detection probabilities (efficiencies) of $D_1$ and $D_2$ ($W_1$ and $W_2$) are the same, also their ratios (reliabilities). In the experiment, in order to make equal optical lengths for the two paths in each MZI, the coherent light needs to be used for the adjustment of the optical lengths.

\section{Numerical result and discussion}
Let us first consider the case without dissipation. In Fig. 3a and 3b, we plot the ratio, $\eta ^{(nb)}$ (the reliability), for Bob does not insert his block (100\% reflection mirrors), and the ratio, $\eta ^{(wb)}$ (the reliability), for Bob blocks his mirrors versus N and M. The result in Fig. 3 is consistent with the result in [16]. Please note $\eta ^{(nb)}$ (no blocks) does not depend on N, and $\eta ^{(wb)}$ (with blocks) always decreases with M for fixed N. Larger M makes larger $\eta ^{(nb)}$ (no blocks), while larger N makes larger $\eta ^{(wb)}$ (with blocks) for fixed M. In principle, as both M and N trend to infinity (the two ratios both go to infinity), detection probability of the photon by $D_1$ or $D_2$ goes to 100\%, and consequently, direct communication(counterfactual for single photon input) between Alice and Bob is achieved.
\begin{figure}
  \centering
  \includegraphics[scale=0.4]{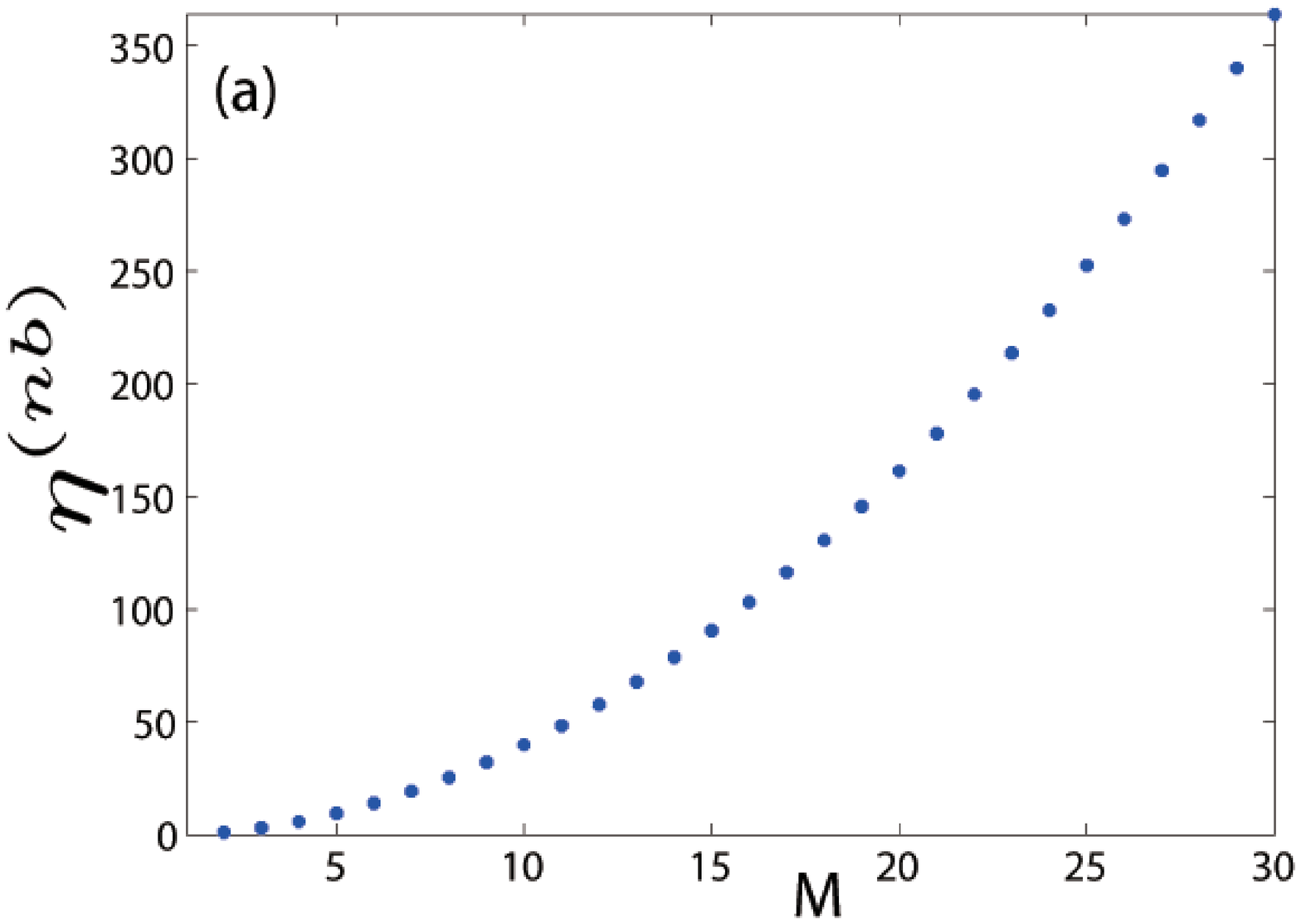}
   \includegraphics[scale=0.45]{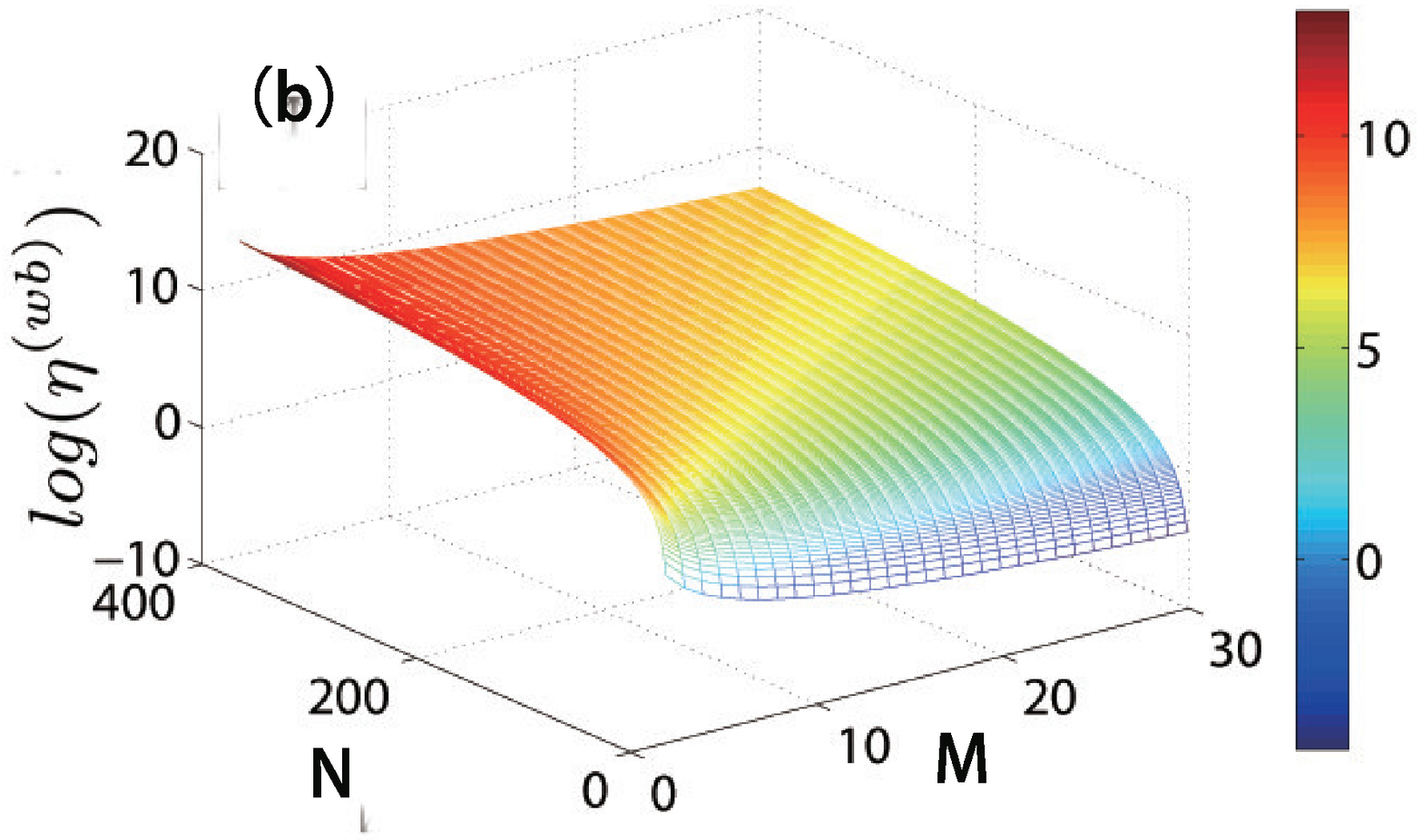}
\caption{The reliability versus N and M without dissipation, (a) Bob does not block the paths ($\eta ^{(nb)}$), (b) Bob blocks the paths $log(\eta ^{(wb)})$, which increase with M. }\label{f3}
\end{figure}

Now let us consider the influence of the dissipation. The energy dissipations in the three path groups  (indicated in Fig. 1(a)) are denoted by  ${\kappa _1}$, ${\kappa _2}$ and ${\kappa _3}$, respectively. In Fig. 4, we plot the reliability, $\eta ^{(wb)}$ (with blocks) versus N and M, respectively, with ${\kappa _2}={\kappa _3}=10^{-4}$ (balanced dissipation) and ${\kappa _1}=3{\kappa _2}$. Please note that the loss of the best quality beam splitter currently available is at the order of ${10^{ - 4}}$  to ${10^{ - 5}}$. When the dissipation is included, we find that $\eta ^{(nb)}$ (no blocks) still increases with M under the balanced dissipation in the inner chain, equivalent to no dissipation (see Fig. 3a).  For no blocks and under the balanced dissipation, the reliability ( $\eta ^{(nb)}$) does not depend on the dissipation (${\kappa _{1,2,3}}$), and the efficiency ($W^{(nb)}_1$) only depends on ${\kappa _3}$. However, the reliability $\eta ^{(wb)}$ increases, and then decreases with N if M larger than a certain value, see Fig. 4, due to the dissipation. Here we ask ourselves,``Can we increase the reliability ($\eta ^{(wb)}$) by some means (not by changing N)?", not by increasing N.
\begin{figure}
  \centering
  \includegraphics[scale=0.4]{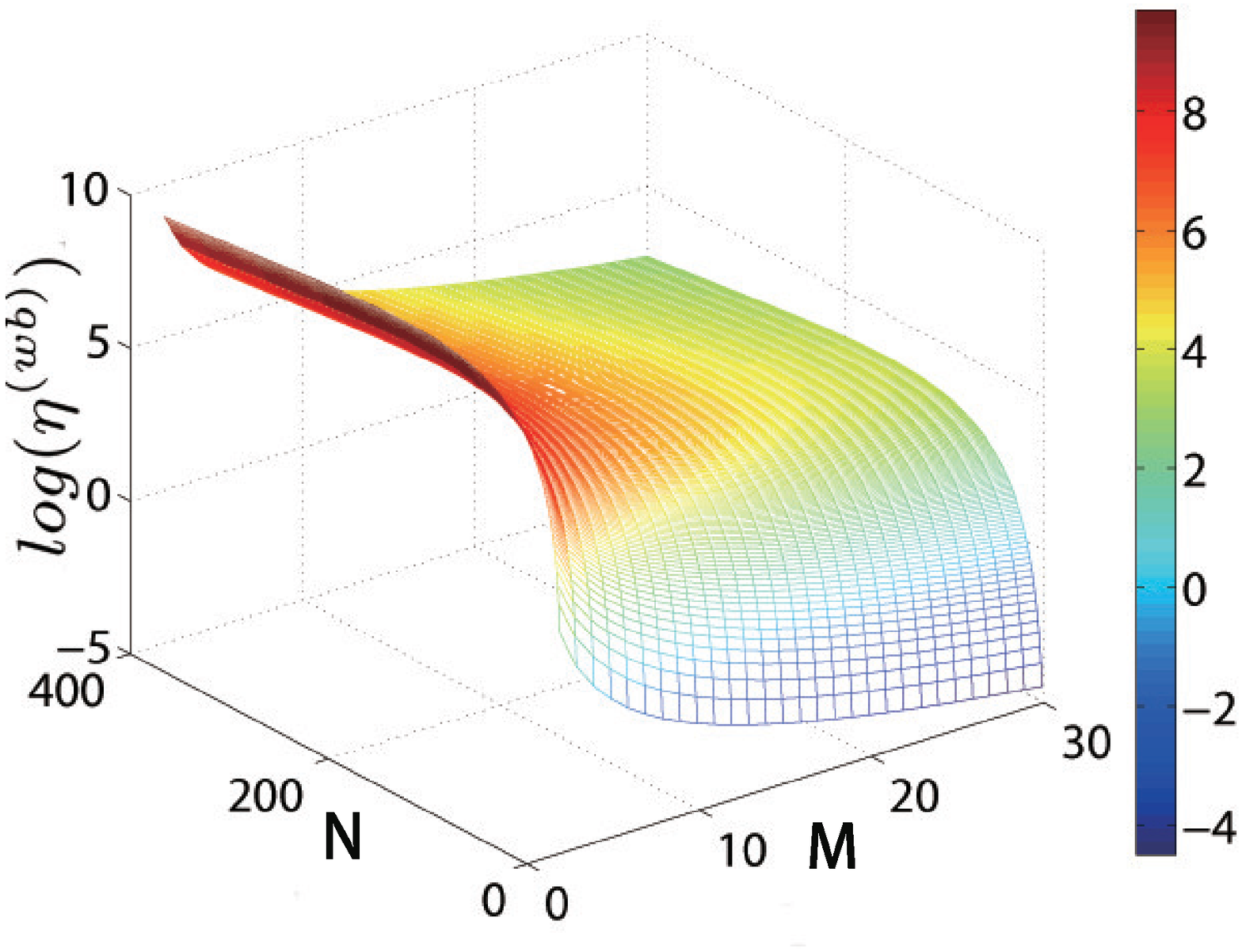}
\caption{The reliability ($log(\eta ^{(wb)})$) versus N and M with dissipation ${\kappa _2}={\kappa _3}=10^{-4}$ and ${\kappa _1}=3\kappa_3$. For lager M, N region, the reliability $\eta ^{(wb)}$ decreases with M and N. }\label{f4}
\end{figure}

\section{The improvement of the reliability ($\eta ^{(wb)}$) by the balanced dissipation with the blocks inserted}

In this section, the balanced dissipation of the inner chain (${\kappa _2} = {\kappa _3}$)  is assumed and Bob inserts the blocks in his paths. Even no dissipation, some photons will be lost due to the blocks for finite N (note no photon loss only for infinity N). Please note the output state at $D_1$ is not the vacuum for finite M, even no dissipation. Some of the photon probability entering the inner chain will back to the outer chain, which results in interference between the paths of the left group (the left paths of the out chain) and the paths of the middle group (left paths of inner chain) at the BSs of the outer chain (each MZI of the outer chain). As discussed above, for no blocks, the use of the balanced dissipation ( ${\kappa _2} = {\kappa _3}$) in the inner chain makes the inner chain equivalent to no dissipation. Can we use the idea of the balanced dissipation in the outer chain to obtain high reliability $\eta ^{(wb)}$ (with blocks), ever $\eta ^{(wb)} \to \infty$? The answer is yes. The interference at these BSs of the outer chain is dependent on ${\kappa _1}$ , that is to say, the interference can be adjusted by  ${\kappa _1}$, and so does the output at $D_1$. When Bob blocks his paths (${\kappa _3}{\text{ = }}1$), the proportion of photon returned back to the outer chain from the inner chain, is  ${\cos ^2}{\theta _N}{\left( {\sqrt {1 - {\kappa _2}} \cos {\theta _N}} \right)^{2(N - 1)}}$, which can be viewed an equivalent dissipation in the paths of the middle group, ${\kappa '_2} = 1 - {\cos ^2}{\theta _N}{\left( {\sqrt {1 - {\kappa _2}} \cos {\theta _N}} \right)^{2(N - 1)}}$. If we introduce a dissipation in the paths of the left group (outer chain),
\begin{equation}\label{kappa1}
  {\kappa_1} = 1 - {\cos ^2}{\theta _N}{\left( {\sqrt {1 - {\kappa _2}} \cos {\theta _N}} \right)^{2(N - 1)}}
\end{equation}
we can achieve a complete interference at the BSs of the MZIs of the outer chain, which results in the vacuum state for the output at $D_1$, and consequently we have $\eta ^{(wb)} \to \infty $  (highest reliability). Please note ${\kappa _2}$  can be zero (no dissipation) in Eq. (10). By adjusting the dissipation, Alice and Bob with finite N and M can have a better communication compared to the case of no dissipation. In Fig. 5, we plot the influence of  $\kappa_1$ and $\kappa_2$ on $\eta ^{(wb)}$ for N=12 and M=6, where we can see $\eta ^{(wb)} \to \infty $ ($W_1=0$, no photon probability for $D_1$) when Eq. (10) is satisfied.

Here we would like to emphasize that $\eta ^{(wb)}$ (no blocks) is not affected by ${\kappa _1}$, if we set the balanced dissipation in the inner chain   ${\kappa _3} = {\kappa _2}$ (including ${\kappa _3} = {\kappa _2} = 0$), because no photon probability from the inner chain back to the outer chain. Therefore, we can manipulate the dissipation ${\kappa _1}$  to maximize $\eta ^{(wb)}$ (with blocks) which has no effect on $\eta ^{(nb)}$ (no blocks). By using the balanced dissipation (for both inner and outer chains), we can improve the communication between Alice and Bob with a few N and M.
\begin{figure}
  \centering
  \includegraphics[scale=0.4]{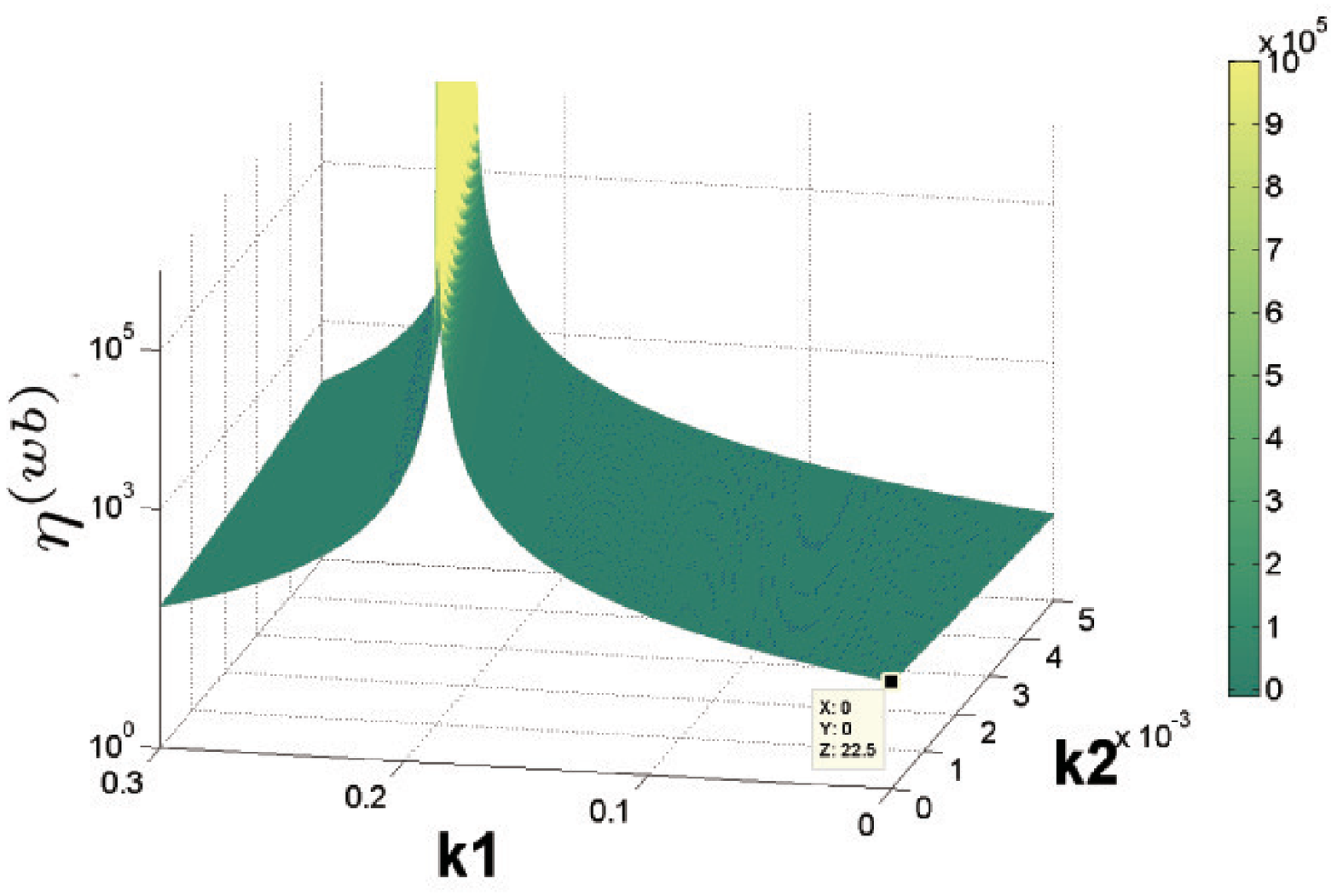}
\caption{The influence of $\kappa_1$ and $\kappa_2$ on $\eta ^{(wb)}$ for N= 12 and M=6.}\label{f5}
\end{figure}

However, the benefit of this method for the reliability improvement is not free. What is the expense? Here we consider the efficiency (the photon probability entering $D_2$), $W^{(wb)}_2$, and the total photon probability in all paths in the transmission channel ($W^{(wb)}_{Tr}$), when Bob inserts the blocks. The efficiency $W^{(wb)}_2$ for N=12 and M=6 is 62\% without the dissipation, ${\kappa _1} = {\kappa _3} = {\kappa _2} = 0$ , which decreases to 36\% with the use of the balanced dissipation (Eq. 10). The balanced dissipation method can make very high reliability ($\eta ^{(wb)} \to \infty $) with the expense of reducing the efficiency ($W^{(wb)}_2$), see Table I.
Now let us consider the total photon probability in the transmission channel, $W^{(wb)}_{Tr}$, which is listed for different combination of M and N with the balanced dissipation method of Eq. (10) and ${\kappa _3} = {\kappa _2} = 0,{\text{ 1}}{{\text{0}}^{ - 4}}$ in Table I, where $W^{(wb)}_{Tr}$ for ${\kappa _1} = {\kappa _2} = {\kappa _3} = 0$  is also listed for comparison. It is clear that the balanced dissipation method can also reduce the photon probability in the transmission channel, which is another benefit of the balanced dissipation method for finite N and M. For large N and M, the balanced dissipation method will greatly reduce the efficiency.\\

\begin{table}
\begin{tabular}{|c|c|c|c|c|c|c|}\hline

\multirow{4}*{M,N} & \multicolumn{1}{p{26mm}|}{\centering  \tiny {No dissipation}\\ \tiny{$W^{(wb)}_1=1-W^{(wb)}_2-W^{(wb)}_{Tr}$}}
& \multicolumn{2}{p{38mm}|}{\centering \tiny{Balanced dissipation,Eq.(10)} \\ \tiny{$W^{(wb)}_1=0$}} \\  \cline{2-4}
                                                & \multicolumn{1}{c|}{\small{$\kappa_1=\kappa_2=\kappa_3=0$}}& \multicolumn{1}{c|}{\small{$\kappa_2=\kappa_3=0$}}& \multicolumn{1}{c|}{\small{$\kappa_2=\kappa_3=10^{-4}$}}\\  \cline{2-4}
& \multicolumn{1}{c|}{$W^{(wb)}_2$\qquad $W^{(wb)}_{Tr}$}& \multicolumn{1}{c|}{$W^{(wb)}_2$\qquad $W^{(wb)}_{Tr}$} & \multicolumn{1}{c|}{$W^{(wb)}_2$\qquad $W^{(wb)}_{Tr}$}  \\   \cline{2-4}
\hline
  6,12     & 0.62   \qquad 0.35 & 0.36  \qquad 0.26 & 0.35  \qquad 0.26\\
 12,12   & 0.37   \qquad 0.54 & 0.10  \qquad 0.27 & 0.10  \qquad 0.27\\
  12,20   & 0.54   \qquad 0.43 & 0.26  \qquad 0.28 & 0.25  \qquad 0.27 \\
  20,30   &0.49    \qquad 0.46 & 0.21  \qquad 0.28 & 0.20  \qquad 0.27 \\
  20,50   & 0.64   \qquad 0.34 & 0.39  \qquad0.25  & 0.36  \qquad 0.23 \\
  30,50   & 0.48   \qquad 0.42 & 0.24  \qquad0.28  & 0.21  \qquad 0.25  \\
  40,100 & 0.63   \qquad 0.35 & 0.38  \qquad 0.25 & 0.26  \qquad 0.20 \\  \cline{2-4}
   \hline
\end{tabular}
\caption{The efficiency (the photon probability entering $D_2$), $W^{(wb)}_2$, and the total photon probability in all paths in the transmission channel ($W^{(wb)}_{Tr}$) for different dissipations and M, N, when Bob inserts the blocks. } \label{tab} \end{table}

\section{Conclusion}
We analyzed the effect of dissipation and finite N and M on the direct communication. We have proposed a balanced dissipation method, Eq. (10) and ${\kappa _2} = {\kappa _3}$, to improve the communication with finite number of BSs. Our derivation is based on operators and is independent of the input states. The results will be the same for different input states: a single photon state and a coherent state. The ratios between the two photon probabilities at $D_1$ and $D_2$ for single photon input are equal to the ratios between the two intensities received by the two detectors. For single photon input, we need to use single photon detectors, while for the coherent state input, we need intensity detectors. In experiments, the coherent state is much easier to produce and to control compared with the single photon state, and can be used to prove, in principle, the protocol. However, the communication is counterfactual for single photon input; it is not for coherent state input.

 \begin{acknowledgements}
This work was supported by National Basic Research Program of China (Grants Nos. 2012CB921601 and 2011CB922203) and National Natural Science Foundation of China (Grants No.11174026 and No.U1330203 ).
\end{acknowledgements}

\begin{widetext}
\appendix
\section{The inner chain}
Let us consider one of the MZIs in the inner chain, for example the first one, see Fig. 6. The inputs are  ${a'_1}^\dag $and  ${a'_0}^\dag$, while the outputs are ${a''_l}^\dag$ and  ${a''_r}^\dag $ with two dissipations ${\kappa _2}$ and ${\kappa _3}$. We can use matrix method to obtain the outputs. The two BSs can be expressed by the same matrix	
$\left[ {\begin{array}{*{20}{c}}
  {\cos \theta }&{ - \sin \theta } \\
  {\sin \theta }&{\cos \theta }
\end{array}} \right]$.
The two dissipations can be expressed by the matrixes\\
  $\left[ {\begin{array}{*{20}{c}}
  1&0 \\
  0&{\sqrt {1 - {\kappa _3}} }
\end{array}} \right]$  and
$\left[ {\begin{array}{*{20}{c}}
  {\sqrt {1 - {\kappa _2}} }&0 \\
  0&1
\end{array}} \right]$.	\\

The relation between the output and the inputs can be written as
\begin{align}
\left( {\begin{array}{*{20}{c}}
  {{a''_l}^\dag } \\
  {{a''_r}^\dag }
\end{array}} \right) &= \left[ {\begin{array}{*{20}{c}}
  {\cos {\theta _N}}&{ - \sin {\theta _N}} \\
  {\sin {\theta _N}}&{\cos {\theta _N}}
\end{array}} \right]\left\{ {\left[ {\begin{array}{*{20}{c}}
  {\sqrt {1 - {\kappa _2}} }&0 \\
  0&\sqrt {1 - {\kappa _3}}
\end{array}} \right]\left[ {\begin{array}{*{20}{c}}
  {\cos {\theta _N}}&{ - \sin {\theta _N}} \\
  {\sin {\theta _N}}&{\cos {\theta _N}}
\end{array}} \right]} \right\}\left( {\begin{array}{*{20}{c}}
  {{a'_1}^\dag } \\
  {{a'_0}^\dag }
\end{array}} \right) \nonumber \\
                               & = \left( {\begin{array}{*{20}{c}}
  {\left( {\sqrt {1 - {\kappa _2}} {{\cos }^2}{\theta _N} - \sqrt {1 - {\kappa _3}} {{\sin }^2}{\theta _N}} \right){a'_1}^\dag } \\
  {\sqrt {1 - {\kappa _2}} {\text{ + }}\sqrt {1 - {\kappa _3}} \sin {\theta _N}\cos {\theta _N}{a'_1}^\dag }
\end{array}} \right).
\end{align}

Equation (3) can also be expressed as,
\begin{equation}
  {a'_1}^\dag  \to (\sqrt {1 - {\kappa _2}} {\cos ^2}{\theta _N} - \sqrt {1 - {\kappa _3}} {\sin ^2}{\theta _N}){a''_l}^\dag  + (\sqrt {1 - {\kappa _2}} {\text{ + }}\sqrt {1 - {\kappa _3}} )\sin {\theta _N}\cos {\theta _N}{a''_r}^\dag
\end{equation}
where we have deleted the vacuum  as it has no contribution to the outputs.

All the MZIs in the inner chain can be derived the same way as Eq. (A1). Consequently, the outputs of the inner chain can be obtained,
\begin{equation}
  \left( {\begin{array}{*{20}{c}}
  {{a'_l}^\dag } \\
  {{a'_r}^\dag }
\end{array}} \right) = \left[ {\begin{array}{*{20}{c}}
  {\cos {\theta _N}}&{ - \sin {\theta _N}} \\
  {\sin {\theta _N}}&{\cos {\theta _N}}
\end{array}} \right]{\left\{ {\left[ {\begin{array}{*{20}{c}}
  {\sqrt {1 - {\kappa _2}} }&0 \\
  0&{\sqrt {1 - {\kappa _3}} }
\end{array}} \right]\left[ {\begin{array}{*{20}{c}}
  {\cos {\theta _N}}&{ - \sin {\theta _N}} \\
  {\sin {\theta _N}}&{\cos {\theta _N}}
\end{array}} \right]} \right\}^{N - 1}}\left( {\begin{array}{*{20}{c}}
  {{a'_1}^\dag } \\
  {{a'_0}^\dag }
\end{array}} \right)
\end{equation}

\begin{figure}
  \centering
  \includegraphics[scale=1]{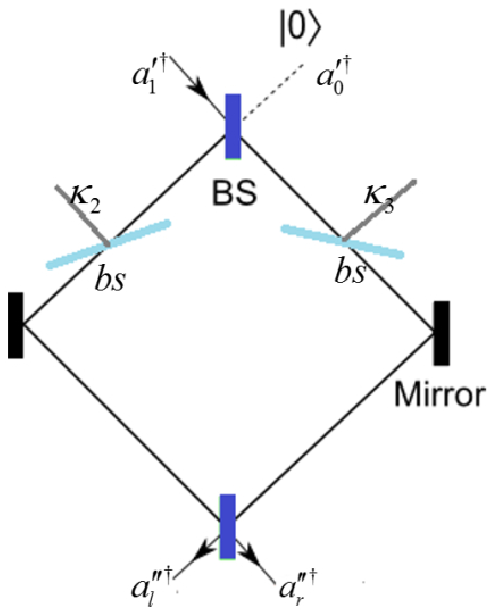}\\
  \caption{MZI}
\end{figure}
Same with (A2), the total transformation can be rewritten as:\\
\begin{equation}
  {a'_1}^\dag  \to {M'_{11}}{a'_l}^\dag  + {M'_{21}}{a'_r}^\dag
\end{equation}
where
\begin{subequations}
\begin{align}
{M'_{11}} = [\begin{array}{*{20}{c}}
  1&0
\end{array}]\left[ {\begin{array}{*{20}{c}}
  {\cos {\theta _N}}&{ - \sin {\theta _N}} \\
  {\sin {\theta _N}}&{\cos {\theta _N}}
\end{array}} \right]{\left\{ {\left[ {\begin{array}{*{20}{c}}
  {\sqrt {1 - {\kappa _2}} }&0 \\
  0&{\sqrt {1 - {\kappa _3}} }
\end{array}} \right]\left[ {\begin{array}{*{20}{c}}
  {\cos {\theta _N}}&{ - \sin {\theta _N}} \\
  {\sin {\theta _N}}&{\cos {\theta _N}}
\end{array}} \right]} \right\}^{N - 1}}\left[ {\begin{array}{*{20}{c}}
  1 \\
  0
\end{array}} \right]        \\
{M'_{21}} = [\begin{array}{*{20}{c}}
  0&1
\end{array}]\left[ {\begin{array}{*{20}{c}}
  {\cos {\theta _N}}&{ - \sin {\theta _N}} \\
  {\sin {\theta _N}}&{\cos {\theta _N}}
\end{array}} \right]{\left\{ {\left[ {\begin{array}{*{20}{c}}
  {\sqrt {1 - {\kappa _2}} }&0 \\
  0&{\sqrt {1 - {\kappa _3}} }
\end{array}} \right]\left[ {\begin{array}{*{20}{c}}
  {\cos {\theta _N}}&{ - \sin {\theta _N}} \\
  {\sin {\theta _N}}&{\cos {\theta _N}}
\end{array}} \right]} \right\}^{N - 1}}\left[ {\begin{array}{*{20}{c}}
  1 \\
  0
\end{array}} \right]
\end{align}
\end{subequations}

\section{The outer chain}
The left output of inner chain will go back to outer chain. The input of the inner chain comes from outer chain together with a vacuum. Hence, the inner chain can be regarded as the right path of the MZIs of the outer chain with dissipation $1 - {M'_{11}}^2$. In this way, the coefficients (${M_1}$ and ${M_2}$) of  $a_R^\dag $ and $a_L^\dag $ in Eq. (3) can be calculated by
\begin{subequations}
\begin{align}
  {M_1} = [\begin{array}{*{20}{c}}
  1&0
\end{array}]\left[ {\begin{array}{*{20}{c}}
  {\cos {\theta _M}}&{ - \sin {\theta _M}} \\
  {\sin {\theta _M}}&{\cos {\theta _M}}
\end{array}} \right]{\left\{ {\left[ {\begin{array}{*{20}{c}}
  {\sqrt {1 - {\kappa _1}} }&0 \\
  0&{{{M'}_{11}}}
\end{array}} \right]\left[ {\begin{array}{*{20}{c}}
  {\cos {\theta _M}}&{ - \sin {\theta _M}} \\
  {\sin {\theta _M}}&{\cos {\theta _M}}
\end{array}} \right]} \right\}^{M - 1}}\left[ {\begin{array}{*{20}{c}}
  1 \\
  0
\end{array}} \right]        \\
{M_2} = [\begin{array}{*{20}{c}}
  0&1
\end{array}]\left[ {\begin{array}{*{20}{c}}
  {\cos {\theta _M}}&{ - \sin {\theta _M}} \\
  {\sin {\theta _M}}&{\cos {\theta _M}}
\end{array}} \right]{\left\{ {\left[ {\begin{array}{*{20}{c}}
  {\sqrt {1 - {\kappa _1}} }&0 \\
  0&{{{M'}_{11}}}
\end{array}} \right]\left[ {\begin{array}{*{20}{c}}
  {\cos {\theta _M}}&{ - \sin {\theta _M}} \\
  {\sin {\theta _M}}&{\cos {\theta _M}}
\end{array}} \right]} \right\}^{M - 1}}\left[ {\begin{array}{*{20}{c}}
  1 \\
  0
\end{array}} \right]
\end{align}
\end{subequations}
The coefficient, ${M_{3i}}$, is one of the two output fields of the i\emph{th} inner chain (see Fig. 1a). Please note ${M_{3i}}$ is proportional the input field of the i\emph{th} inner chain, which is one of the two output fields of the i\emph{th} MZI of the outer chain,${M_{i(inner)}}{a'_r}^\dag $ . The output of the i\emph{th} MZI of the outer chain can be obtained with the same method in Appendix A with  ${\theta _N}$ replaced by ${\theta _M}$, and the dissipation matrix
$\left[ {\begin{array}{*{20}{c}}
  {\sqrt {1 - {\kappa _2}} }&0 \\
  0&{\sqrt {1 - {\kappa _3}} }
\end{array}} \right]$
 replaced by
  $ \left[ {\begin{array}{*{20}{c}}
  {\sqrt {1 - {\kappa _1}} }&0 \\
  0&{{{M'}_{11}}}
\end{array}} \right] $,
so that we have
\begin{align}
{M_{i(inner)}} = [\begin{array}{*{20}{c}}
  0&1
\end{array}]\left[ {\begin{array}{*{20}{c}}
  {\cos {\theta _M}}&{ - \sin {\theta _M}} \\
  {\sin {\theta _M}}&{\cos {\theta _M}}
\end{array}} \right]{\left\{ {\left[ {\begin{array}{*{20}{c}}
  {\sqrt {1 - {\kappa _1}} }&0 \\
  0&{{{M'}_{11}}}
\end{array}} \right]\left[ {\begin{array}{*{20}{c}}
  {\cos {\theta _M}}&{ - \sin {\theta _M}} \\
  {\sin {\theta _M}}&{\cos {\theta _M}}
\end{array}} \right]} \right\}^{i - 1}}\left[ {\begin{array}{*{20}{c}}
  1 \\
  0
\end{array}} \right]
\end{align}
with the coefficient ${M_{i(inner)}}$  in hands, the coefficient of output operator to $D_{3i}$, ${M_{3i}}$, can be given
\begin{align}
{M_{3i}} = [\begin{array}{*{20}{c}}
  0&1
\end{array}]\left[ {\begin{array}{*{20}{c}}
  {\cos {\theta _N}}&{ - \sin {\theta _N}} \\
  {\sin {\theta _N}}&{\cos {\theta _N}}
\end{array}} \right]{\left\{ {\left[ {\begin{array}{*{20}{c}}
  {\sqrt {1 - {\kappa _2}} }&0 \\
  0&{\sqrt {1 - {\kappa _3}} }
\end{array}} \right]\left[ {\begin{array}{*{20}{c}}
  {\cos {\theta _N}}&{ - \sin {\theta _N}} \\
  {\sin {\theta _N}}&{\cos {\theta _N}}
\end{array}} \right]} \right\}^{N - 1}}
\left[ {\begin{array}{*{20}{c}}
  {{M_{i(inner)}}} \\
  0
\end{array}} \right]
 \end{align}
\end{widetext}

\end{document}